\UseRawInputEncoding
\documentclass[12pt,aps,floats,showpacs,amssymb,tightenlines]{revtex4}

\usepackage{color}
\usepackage{graphicx}
\usepackage{amsmath}
\usepackage{amsfonts}
\usepackage{amscd}
\usepackage{epsfig}
\usepackage{amssymb}
\usepackage{tabularx}
\usepackage{float}

\begin{document}

\title{The Linet - Tian metrics are a restricted set of Krasi\'nski's
solutions of Einstein's field equations for a rotating perfect fluid.}
\author{Reinaldo J. Gleiser} \email{gleiser@fis.uncor.edu}

\affiliation{Instituto de F\'{\i}sica Enrique Gaviola and FAMAF,
Universidad Nacional de C\'ordoba, Ciudad Universitaria, (5000)
C\'ordoba, Argentina}

\begin{abstract}

In this note we show that the Linet - Tian family of solutions of
the vacuum Einstein equations with a cosmological constant are a
restricted set of the solutions of the Einstein field equations for
a rotating perfect fluid previously found by A. Krasi\'nski.

\end{abstract}

\pacs{04.20.Jb}

\maketitle

\section{Introduction}

There is an interesting family of solutions of the vacuum Einstein
field equations with a cosmological constant $\Lambda$, that can be
positive or negative, found independently by Linet \cite{linet} and
Tian \cite{tian}. The solutions are static, and contain also two
other orthogonal Killing vectors. They may be written in the form
\cite{bla1},
\begin{eqnarray}\label{1}
    ds^2& = & -y^{1/3+ p_1/2} (1-\Lambda y)^{1/3- p_1/2}C_1 dt^2+\frac{1}{3 y (1-\Lambda y)
    } dy^2 \\
    & &+y^{1/3+ p_2/2} (1-\Lambda y)^{1/3- p_2/2} C_2dz^2+y^{1/3+ p_3/2} (1-\Lambda y)^{1/3- p_3/2}
    C_3 d\phi^2 \nonumber
\end{eqnarray}
where the $C_i$ are arbitrary constants, and, for $\Lambda > 0$, $y$
is restricted to $0 < y < 1/\Lambda$, while for $\Lambda < 0$ the
range of $y$ is $0 \leq y \leq \infty$. They satisfy Einstein's
equations \cite{bla2},
\begin{equation}\label{2}
    G_{\mu\nu} = -\Lambda g_{\mu\nu}
\end{equation}
provided the parameters $p_i$ satisfy the relations,
\begin{eqnarray}
\label{3}
  p_1+p_2+p_3 &=& 0 \\
  p_1^2+p_2^2+p_3^2 &=& \frac{8}{3} \nonumber
\end{eqnarray}
and this, in turn implies that the allowed values of the parameters
$p_i$ are restricted to the range,
\begin{equation}\label{3a}
  -4/3 \leq  p_i \leq 4/3
\end{equation}
and all values in that range are allowed. Clearly, (\ref{3}) is
still satisfied if we change every $p_i$ to $-p_i$, so that, given a
solution of (\ref{3}), we get a new solution by simply changing the
sign of every $p_i$. However, this new solution is diffeomorphic to
the original one, as can be seen by changing coordinates in
(\ref{1}), in accordance with,
\begin{equation}\label{3b}
    y=(1-\Lambda z)/\Lambda
\end{equation}
which changes (\ref{1}) to,
\begin{eqnarray}\label{1a}
    ds^2& = & -z^{1/3- p_1/2} (1-\Lambda y)^{1/3+ p_1/2}\widetilde{C}_1 dt^2+\frac{1}{3 y (1-\Lambda y)
    } dy^2 \\
    & &+y^{1/3- p_2/2} (1-\Lambda y)^{1/3+ p_2/2} \widetilde{C}_2dz^2+y^{1/3- p_3/2} (1-\Lambda y)^{1/3+ p_3/2}
    \widetilde{C}_3 d\phi^2 \nonumber
\end{eqnarray}
where $\widetilde{C}_i=C_i/\Lambda^{p_i}$, which is identical to
(\ref{1}), up to a rescaling of the remaining coordinates, and the
change $p_i \to -p_i$. Similarly, we get the same solution, up to a
rescaling of appropriate coordinates, by exchanging $p_2$ and $p_3$.
Finally we notice that we may solve (\ref{3}) for $p_2$ and $p_3$ to
get,
\begin{eqnarray}\label{3c}
    p_2 & =& -\dfrac{p_1}{2}+\epsilon \dfrac{\sqrt{48-27 p_1{}^2}}{6}  \\
    p_3 &= &-\dfrac{p_1}{2}-\epsilon \dfrac{\sqrt{48-27 p_1{}^2}}{6} \nonumber
\end{eqnarray}
where $\epsilon=\pm 1$. This can also be seen as a proof of
(\ref{3a}). As just discussed, we will set $\epsilon=1$, without
loss of generality. These properties will be important in the
discussions that follows.
\\

The properties and applications of the Linet - Tian metrics have
been the subject of a number of studies. A recent review of these
and similar types of metrics has been recently presented in
\cite{bronnikov}. It apparently has escaped this review, as well as
most the research papers centering on these type of metrics, that
the Linet - Tian solutions are contained, as a restricted set, in a
large family of solutions of the vacuum Einstein equations with a
cosmological constant, previously found by A. Krasi\'nski
\cite{krasinski1}, \cite{krasinski2}. These metrics can be obtained
from the following Ansatz \cite{simply}. If one writes the metrics
in the form,
\begin{eqnarray}
\label{4}
 ds^2 &=& \frac{1}{v^{2/3}}dx_0{}^2+2\frac{x_2}{v^{2/3}}dx_0 dx_1
 +\frac{x_2{}^2-V}{v^{2/3}} dx_1{}^2 \\
   & & -\frac{J^2}{s v^2} \exp\left(-\int{\frac{x_2}{V} dx_2}\right) dx_2{}^2
   -\frac{V}{s v^{2/3}} \exp\left(-\int{\frac{x_2}{V} dx_2}\right) dx_3{}^2\nonumber
\end{eqnarray}
where $V=V(x_2)$, $v=v(x_2)$, and $J$ and $s$ are constants, then, a
necessary condition for the Einstein equations (\ref{2}) to be
satisfied is that $V$ is a solution of,
\begin{equation}\label{5}
    \frac{d^2V}{d x_2{}^2}-2=0
\end{equation}
and $v$ is a solution of,
\begin{eqnarray}
\label{6}
  \frac{d^2v}{d x_2{}^2}&=&  \frac{1}{V}\left(\frac{dV}{dx_2}-x_2\right) \frac{dv}{dx_2}\\
   & & +\frac{3}{4V^2}\left(V \frac{d^2V}{dx_2{}^2}-V+x_2 \frac{dV}{dx_2}
   -\left(\frac{dV}{dx_2}\right)^2\right) v \nonumber
\end{eqnarray}

Notice that (\ref{5}) and (\ref{6}) are independent of $\Lambda$,
and, in fact, of the signature of (\ref{4}). The general (real)
solution of (\ref{5}) can be written in form,
\begin{equation}\label{8}
   V(x_2)=(x_2-q_0)(x_2-p_0)
\end{equation}
where $p_0$, and $q_0$, are constants, and we have three
possibilities, namely, $p_0$, and $q_0$, are real and distinct,
$p_0=q_0$, (both real), and $p_0=q_0^{*}$, i.e., complex conjugate
of each other.

The Linet-Tian metrics (\ref{1}) have three orthogonal Killing
vectors, while this, in principle, is not the case for the
Krasi\'nski metrics (\ref{4}). The terms in question are,
\begin{eqnarray}\label{8a}
  d \sigma^2 & =&  \frac{1}{v^{2/3}}dx_0{}^2+2\frac{x_2}{v^{2/3}}dx_0 dx_1
 +\frac{x_2{}^2-V}{v^{2/3}} dx_1{}^2 \\
 & & =
 \frac{1}{v^{2/3}}\left(dx_0{}^2+2 x_2 dx_0 dx_1
 +\left((p_0+q_0) x_2-p_0q_0\right) dx_1{}^2 \right) \nonumber
\end{eqnarray}
implying that $\partial_{x_0}$, and $\partial_{x_1}$ are not
orthogonal. We, therefore, consider a (linear) change of coordinate
basis of the form,
\begin{eqnarray}\label{8b}
  x_0 & = & a_1 y_0+a_2 y_1  \\
 x_1 &= & b_1 y_0 +b_2 y_1 \;\;,
   \nonumber
\end{eqnarray}
and find that the conditions for the orthogonality of
$\partial_{y_0}$, and $\partial_{y_1}$ are,
\begin{equation}\label{8c}
  b_1 b_2 - a_1 a_2 p_0 q_0 = 0 \;,
  \end{equation}
and either,
\begin{equation}\label{8d}
   a_2 q_0 + b_2 = 0 \;,
  \end{equation}
or,
\begin{equation}\label{8e}
   a_2 p_0 + b_2 = 0 \;.
\end{equation}
Solving (\ref{8c}) for $b_1$, and (\ref{8d}) for $a_2$, we get,
\begin{equation}\label{8f}
  d \sigma^2 =  \frac{a_1{}^2
  (x_2-p_0)(q_0-p_0)}{p_0{}^2v^{2/3}}dy_0{}^2
 +
 \frac{b_2{}^2
  (q_0-x_2)(q_0-p_0)}{ v^{2/3}}dy_1{}^2 \; ,
\end{equation}
while solving (\ref{8e}) for $a_2$, and replacing in (\ref{8a}), we
find,
\begin{equation}\label{8g}
  d \sigma^2 =  \frac{a_1{}^2
  (q_0-x_2)(q_0-p_0)}{q_0{}^2v^{2/3}}dy_0{}^2
 +
 \frac{b_2{}^2
  (x_2-p_0)(q_0-p_0)}{ v^{2/3}}dy_1{}^2\;.
\end{equation}

We first notice that since $a_1$ and $b_2$ are arbitrary, (\ref{8f})
and (\ref{8g}) are equivalent up to irrelevant changes of names.
Next we see that for complex $q_0$ and $p_0$ the transformation
(\ref{8b}) leads to complex coefficients in $d\sigma^2$, while for
$p_0=q_0$ the transformation is singular. The only acceptable case
is then for $q_0$ and $p_0$ real and distinct. In what follows we
assume, without loss of generality, $q_0 > p_0$. The Krasi\'nski
metric (\ref{4}) then takes the form,
\begin{eqnarray}
\label{4a}
 ds^2 &=& \frac{
  (x_2-p_0)}{v^{2/3}}dy_0{}^2
 +
 \frac{
  (q_0-x_2)}{ v^{2/3}}dy_1{}^2  \\
   & & -\frac{J^2}{s v^2} \exp\left(-\int{\frac{x_2}{V} dx_2}\right) dx_2{}^2
   -\frac{V}{s v^{2/3}} \exp\left(-\int{\frac{x_2}{V} dx_2}\right) dx_3{}^2\nonumber
\end{eqnarray}
where, without loss of generality, we have have chosen (\ref{8f}),
and assigned values to $a_1$, and $b_2$ so as to simplify the
resulting expressions. With this restriction we still have to
consider three separate cases, namely, $x_2 \geq q_0$, $q_0\geq x_2
\geq p_0$, and $p_0 \geq x_2$. These are analyzed in what follows.

\section{The case $x_2 \geq q_0$.}

\subsection{The form of the metric.}

To continue our analysis of (\ref{4}) we notice that for $x_2 \geq
q_0$,
\begin{eqnarray}
\label{9}
    \exp\left(-\int{\frac{x_2}{V}dx_2}\right)&=&
 \exp\left(-\int{\frac{x_2}{(x_2-p_0)(x_2-q_0)}}\right) \\
& = & C
(x_2-p_0)^{\frac{p_0}{q_0-p_0}}(x_2-q_0)^{\frac{-q_0}{q_0-p_0}}
\nonumber
\end{eqnarray}
where $C$ is a constant. Since the left hand side of (\ref{9}) is
real and positive, we may set $C=1$.

Again for $x_2 \geq q_0$, we may write the solution of (\ref{6}) in
the form,
\begin{equation}
\label{10}
 v \left( x_{{2}} \right)= \frac{ \left( x_{{2}}-p_0 \right)^{{\frac
 {p_0-2q_0}{2(p_0-q_0)}}-{\frac {\sqrt {{{q_0 }}^{2}-p_0\,{
q_0}+{p_0}^{2}}}{2(p_0-q_0)}}}
 \left( x_{{2}}-q_0 \right) ^{{\frac {2 p_0-q_0}{2(p_0-q_0)}}-{\frac {\sqrt {{q_0}^{2}
-p_0\,q_0+{p_0}^{2}}}{2(p_0-q_0)}}}} { \left( Q
 \left( x_{{2}}-q_0 \right) ^{
 {\frac {\sqrt {{q_0}^{2}-{
p_0}\,q_0+{p_0}^{2}}}{p_0-q_0}}}+P \left( x_{{2}}-{  p_0} \right)
^{{\frac {\sqrt {{q_0}^{2}-p_0\,q_0+{{  p_0}}^{2}}}{p_0-{q_0}}}}
\right)}
\end{equation}
where $P$ and $Q$ are real constants. In what follows we restrict to
$P>0$, and $Q>0$, to avoid singularities in $V(x_2)$.

Replacing (\ref{8}), (\ref{9}), and (\ref{10}) in (\ref{4}) we find
that the full set of Einstein's equations (\ref{2}) is satisfied if
we impose,
\begin{equation}\label{10a}
   \Lambda = -\frac{s Q P (q_0^2-q_0p_0+p_0^2)}{3 J^2}
\end{equation}

We remark again that equations (\ref{2})  are satisfied
independently of the signature assigned to (\ref{4}), or the
particular signs of $J^2$ or $s$, and, therefore, (\ref{4}) provides
a (possibly large) family of solutions of (\ref{2}). At this point
it is convenient to go back to the ``diagonal'' form (\ref{4a}).
This restricted set of Krasi\'nski's metrics provides solutions of
(\ref{2}) with three commuting Killing vectors ($\partial_{y_0}$,
$\partial_{y_1}$, and $\partial_{x_3}$), but we still need to fix
the signature of the metrics. In the case of $\Lambda
> 0$, if we assume $J^2 >0$, from (\ref{10a}) we must take $s<0$.
Without loss of generality we may take $s=-1$, and this makes the
metric (\ref{4a}) static and with signature $(-,+,+,+)$, and, in
accordance with (\ref{10a}), corresponding to $\Lambda >0$. In more
detail, the ``diagonal'' metric is then given by,
\begin{eqnarray}\label{12}
   ds^2 & = & -\frac{(x_2-q_0)}{v^{3/2}} dy_1{}^2
    +\frac{ J^2
   (x_2-p_0)^{\frac{p_0}{q_0-p_0}}(x_2-q_0)^{-\frac{q_0}{q_0-p_0}}}{
   v^2} dx_2{}^2 \\
   &&
   +\frac{
   (x_2-p_0)^{\frac{-q_0}{q_0-p_0}}(x_2-q_0)^{-\frac{p_0}{q_0-p_0}}}{
   v^{2/3}} dx_3{}^2
   +\frac{(x_2-p_0)}{v^{3/2}} dy_0{}^2 \ \nonumber
\end{eqnarray}
But, the Linet and Tian analysis shows that the static solutions of
(\ref{2}) with three orthogonal commuting Killing vectors are
unique, up to diffeomorphisms. Therefore, the Linet - Tian metrics
and Kransinski's metrics should be related by a coordinate
transformation. That this is the case is shown explicitly in the
next Section.

\subsection{A coordinate transformation.}

We consider again (\ref{1}) and (\ref{12}) and a coordinate
coordinate change of the form $y = y(x_2)$ that would take (\ref{1})
into (\ref{12}). Under this change we should have,
\begin{equation}\label{14}
   \frac{1}{3  y (1- \Lambda y)}\left(\frac{dy}{dx_2}\right)^2 = \frac{ J^2
   (x_2-p_0)^{\frac{p_0}{q_0-p_0}}(x_2-q_0)^{-\frac{q_0}{q_0-p_0}}}{
   v(x_2)^2}
\end{equation}
or,
\begin{equation}\label{15}
   \left[\frac{1}{3  y (1-\Lambda y)}\right]^{1/2}\left(\frac{dy}{dx_2}\right) = \left[\frac{ J^2
   (x_2-p_0)^{\frac{p_0}{q_0-p_0}}(x_2-q_0)^{-\frac{q_0}{q_0-p_0}}}{
   v(x_2)^2}\right]^{1/2}
\end{equation}

This can be integrated to,
\begin{eqnarray}\label{16}
 & & \ln\left(1-2 \Lambda y  -2 i
\sqrt{\Lambda y}\sqrt{1-\Lambda y}\right) \\
 & = &
 \ln\left[\frac{\sqrt{Q}(x_2-q_0)^{\frac{\sqrt{q_0^2+q_0p_0+p_0^2}}{2(p_0-q_0)}}
 -i\sqrt{P}(x_2-p_0)^{\frac{\sqrt{q_0^2+q_0p_0+p_0^2}}{2(p_0-q_0)}}
 }{\sqrt{Q}(x_2-q_0)^{\frac{\sqrt{q_0^2+q_0p_0+p_0^2}}{2(p_0-q_0)}}
 +i\sqrt{P}(x_2-p_0)^{\frac{\sqrt{q_0^2+q_0p_0+p_0^2}}{2(p_0-q_0)}}}\right]
 \nonumber
\end{eqnarray}
where we have used (\ref{10}) to eliminate $\Lambda$. Eq. (\ref{16})
can be solved for $y$.
\begin{equation}\label{17}
     y=\frac{P \left( x_{{2}}-p_0 \right) ^{{\frac {\sqrt {{q_0}^{2}-q_0 p_0+{p_0}^{2}}}{-q_0+p_0}}}}
 {\left[ Q \left(x_2-q_0 \right) ^{{\frac {\sqrt {{q_0}
^{2}-q_0 p_0+{p_0}^{2}}}{-q_0+p_0}}}+P
 \left( x_{{2}}-p_0 \right) ^{{\frac {\sqrt {{q_0}^{2}-q_0
 p_0+{p_0}^{2}}}{-q_0+p_0}}}\right] \Lambda
 }
\end{equation}

The range of $x_2$ is $\left(q_0 \leq x_2 \leq \infty \right)$, and,
in accordance with (\ref{17}), we have,
\begin{equation}\label{17a}
    \frac{1}{\Lambda} \geq y \geq
\frac{P}{\Lambda(P+Q)}  \;\;,  \mbox{  for  }\;\; q_0 \leq x_2 \leq
\infty .
\end{equation}

We must remark that although there is a sign ambiguity in
(\ref{15}), we only need that (\ref{14}) be satisfied, and one can
check that (\ref{17}) satisfies this requirement.

Next we consider the coefficient of $dy_1{}^2$ in (\ref{12}). We
have,
\begin{equation}\label{18}
    -\frac{x_2-q_0}{v^{2/3}}= - \frac{\left( x_{{2}}-q_0 \right) ^{{\frac {p_0-2 q_0
+\sqrt {{q_0}^{2}-q_0\,p_0+{p_0}^{2}}}{3(-q_0+p_0)}}} \left(
x_{{2}}-p_0 \right) ^{{\frac {2 q_0-p_0+\sqrt
{{q_0}^{2}-q_0\,p_0+{p_0}^{2}}}{3(- q_0+p_0)}}}} { \left( Q \left(
x_{{2}}-q_0 \right) ^{{ \frac {\sqrt {{q_0}^{2}-q_0\,p_0+
p_0{}^{2}}}{-q_0 + p_0}}}+P \left( x_{{2}}-p_0 \right) ^{{\frac
{\sqrt {{q_0}^{2}-q_0 p_0+{p_0}^{2}}}{-q_0+p_0 }}} \right) ^{2/3}}
\end{equation}

On the other hand, going back to (\ref{1}), if we consider the
coefficient of $dt^2$, and change variables in accordance with
(\ref{17}), to get,
\begin{eqnarray}\label{19}
    & & -C_1 y^{\frac{1}{3}+\frac{p_1}{2}} (1-\Lambda y)^{\frac{1}{3}-\frac{p_1}{2}}
    = \\
&& - \frac{ C_1 P^{\frac{1}{3}+\frac{p_1}{2}}  \left( x_{{2}}-{p_0}
\right) ^{{\frac {\sqrt {{{q_0}}^{2} -{q_0}{ p_0}+{{p_0}}^{2}}
\left( 2+3 {p_1} \right) }{6(-{  q_0}+{ p_0})}}} \left(
x_{{2}}-{q_0} \right) ^{{ \frac {\sqrt {{q_0}^{2}-q_0 p_0+{p_0}^{2}}
 \left( 2-3 {p_1} \right) }{6(-q_0+p_0)}}}}
 { \Lambda^{\frac{1}{3}+\frac{p_1}{2}}Q^{-\frac{1}{3}+\frac{p_1}{2}} \left( Q
 \left( x_{{2}}-q_0 \right) ^{{\frac {\sqrt {{q_0}^{2}-{
q_0}\,p_0+{p_0}^{2}}}{-q_0+p_0}}}+P \left( x_{{2}}- p_0 \right)
^{{\frac {\sqrt {{q_0}^{2}-q_0\,p_0+{{ p_0}}^{2}}}{-{q_0}+p_0}}}
\right) ^{2/3}} \nonumber
\end{eqnarray}

We then have that (\ref{19}) will be equal to (\ref{18}) if we set,
\begin{equation}\label{20}
    C_1=\Lambda^{\frac{1}{3}+\frac{p_1}{2}}
    P^{-\frac{1}{3}-\frac{p_1}{2}} Q^{-\frac{1}{3}+\frac{p_1}{2}}
\end{equation}
and,
\begin{equation}\label{21}
    p_1=\frac{2(2 q_0-p_0)}{3\sqrt{p_0{}^2-p_0q_0+q_0{}^2}}
\end{equation}

Similarly, for the coefficient of $dy_0{}^2$ in (\ref{12}) we have,
\begin{equation}\label{22}
    \frac{x_2-p_0}{v^{2/3}}=  \frac{\left( x_{{2}}-q_0 \right) ^{{\frac {q_0-2 p_0
+\sqrt {{q_0}^{2}-q_0\,p_0+{p_0}^{2}}}{3(-q_0+p_0)}}} \left(
x_{{2}}-p_0 \right) ^{{\frac {2 p_0-q_0+\sqrt
{{q_0}^{2}-q_0\,p_0+{p_0}^{2}}}{3(-q_0+p_0)}}}} { \left( Q \left(
x_{{2}}-q_0 \right) ^{{ \frac {\sqrt {{q_0}^{2}-q_0\,p_0+
p_0{}^{2}}}{-q_0 + p_0}}}+P \left( x_{{2}}-p_0 \right) ^{{\frac
{\sqrt {{q_0}^{2}-q_0 p_0+{p_0}^{2}}}{-q_0+p_0 }}} \right) ^{2/3}}
\end{equation}
while for the coefficient of $dz^2$ in (\ref{1}) we have,
\begin{eqnarray}\label{23}
    & & C_2 y^{\frac{1}{3}+\frac{p_2}{2}} (1-\Lambda y)^{\frac{1}{3}-\frac{p_2}{2}}
    = \\
&&  \frac{ C_2 P^{\frac{1}{3}+\frac{p_2}{2}}  \left( x_{{2}}-{p_0}
\right) ^{{\frac {\sqrt {{{q_0}}^{2} -{q_0}{ p_0}+{{p_0}}^{2}}
\left( 2+3 {p_2} \right) }{6(-{  q_0}+{ p_0})}}} \left(
x_{{2}}-{q_0} \right) ^{{ \frac {\sqrt {{q_0}^{2}-q_0 p_0+{p_0}^{2}}
 \left( 2-3 {p_2} \right) }{6(-q_0+p_0)}}}}
 { \Lambda^{\frac{1}{3}+\frac{p_1}{2}}Q^{-\frac{1}{3}+\frac{p_1}{2}} \left( Q
 \left( x_{{2}}-q_0 \right) ^{{\frac {\sqrt {{q_0}^{2}-{
q_0}\,p_0+{p_0}^{2}}}{-q_0+p_0}}}+P \left( x_{{2}}- p_0 \right)
^{{\frac {\sqrt {{q_0}^{2}-q_0\,p_0+{{ p_0}}^{2}}}{-{q_0}+p_0}}}
\right) ^{2/3}} \nonumber
\end{eqnarray}
and we have equality of (\ref{23}) and (\ref{22}) imposing,
\begin{equation}\label{24}
    C_2=\Lambda^{\frac{1}{3}+\frac{p_2}{2}}
    P^{-\frac{1}{3}-\frac{p_2}{2}} Q^{-\frac{1}{3}+\frac{p_2}{2}}
\end{equation}
and,
\begin{equation}\label{25}
    p_2=\frac{2(2 p_0-q_0)}{3\sqrt{p_0{}^2-p_0q_0+q_0{}^2}}
\end{equation}

Finally, and using the same similar procedure, for the coefficient
of $d\phi^2$ we find,
\begin{equation}\label{26}
    C_3=\Lambda^{\frac{1}{3}+\frac{p_3}{2}}
    P^{-\frac{1}{3}-\frac{p_3}{2}} Q^{-\frac{1}{3}+\frac{p_3}{2}}
\end{equation}
and,
\begin{equation}\label{27}
    p_3=-\frac{2(p_0+q_0)}{3\sqrt{p_0{}^2-p_0q_0+q_0{}^2}}
\end{equation}

We can immediately check that the $p_i$ in (\ref{21}), (\ref{25}),
and (\ref{27}) satisfy (\ref{3}). Moreover, considering again
(\ref{21}), (\ref{25}), and (\ref{27}) we have three different
cases: $p_0>0$, $p_0=0$, and $p_0<0$. In the case $p_0 >0$ we may
set $p_0=1$, since the $p_i$ depend only on the ratio $q_0/p_0$, and
then considering all $q_0 > p_0$ we find for the $p_i$ the ranges,
\begin{eqnarray}
\label{28}
  \frac{2}{3} &\leq &  p_1 \leq \frac{4}{3} \nonumber\\
  \frac{2}{3} &\geq &  p_2 \geq -\frac{2}{3}  \\
 - \frac{4}{3} &\leq &  p_3 \leq -\frac{2}{3} \nonumber
\end{eqnarray}

Similarly, for $p_0 < 0$ we have,
\begin{eqnarray}
\label{29}
  -\frac{2}{3} &\leq &  p_1 \leq \frac{4}{3} \nonumber\\
  -\frac{2}{3} &\geq &  p_2 \geq -\frac{2}{3}  \\
 + \frac{4}{3} &\leq &  p_3 \leq -\frac{2}{3} \nonumber
\end{eqnarray}

The case $p_0=0$ corresponds just to $p_1=4/3$, $p_2=-2/3$, and
$p_3=-2/3$.

\section{Summary of results.}

Eqs. (\ref{28}) and (\ref{29}), together with the properties already
discussed of the $p_i$ imply that the full relevant range of values
of the $p_i$ are covered with appropriate choices of $p_0$, and
$q_0$. Considering the range of $y$, we notice that once the $p_i$
are fixed, the only relevant quantity is $\Lambda$, since the $C_i$
in (\ref{1}) represent only rescalings of the coordinates. But
according to (\ref{10a}), fixing $\Lambda$ we still have a large
freedom in the choices of $P$, $Q$, and $J^2$. In accordance with
(\ref{17a}), this freedom can be used to cover essentially the full
range of $y$ in (\ref{1}), with the exception of the singular point
$y=0$, so that we have shown that the Linet-Tian metrics are
effectively a restricted set of the Krasi\'nski metrics (\ref{4}),
in the case $x_2 > q_0$. But essentially similar derivations show
that this is also the case for $q_0 \geq x_2 \geq p_0$, and $
p_0\geq x_2$. In the first case the full range $0 \leq y \leq
1/\Lambda$ is covered, while in the second the point $y= 1/\Lambda$
is excluded. This, in turn, shows that the three possible ranges of
$x_2$ in (\ref{4}) correspond, up to isometries, to the same
solution, and also completes the proof of the equivalence of the
metrics (\ref{1}) and (\ref{4}) in the cases where we take $p_0$ and
$q_0$ in (\ref{8}) as real and distinct.

\section*{Acknowledgments}

I am grateful to A. Krasi\'nski for bringing to my attention the
references to his work and its possible relation to the Linet - Tian
metrics, and also for his comments on this note.


\begin{thebibliography}{99}
\bibitem{simply} I have, without loss of generality, and for convenience
 in the derivations that follow, simplified the
expression of the metric (\ref{4}), as compared to the original form
in \cite{krasinski2}.
\end{thebibliography}
\end{document}